\documentclass[aps,prb,floats,twocolumn,floatfix,showpacs]{revtex4}
\usepackage{epsf,graphicx}
\usepackage{amssymb}
\usepackage{amsmath}
\usepackage{eepic}

\newlength{\piclen}
\begin{document}

\title{From $d$- to $p$-wave pairing in
the $t$-$t'$ Hubbard model at zero temperature}

\author{R.\ Arita and  K.\ Held}
\affiliation{Max-Planck-Institut f\"ur
Festk\"orperforschung, 70569 Stuttgart, Germany}

 \date{\today}

\begin{abstract}
We develop a DCA(PQMC) algorithm which employs the projective quantum 
Monte Carlo (PQMC) method for solving the equations of the dynamical 
cluster approximation (DCA) at zero temperature, and apply it for 
studying pair susceptibilities of the two-dimensional Hubbard-model 
with next-nearest neighbor hopping. In particular, we identify which 
pairing symmetry is dominant in the $U$-$n$ parameter space ($U$: repulsive 
Coulomb interaction; $n$: electron density). We find that $p_{x+y}$- 
($d_{x^2-y^2}$-) wave is dominant among triplet (singlet) pairings -at 
least for $0.3< n <0.8$ and $U\leq 4t$. The crossover between
$d_{x^2-y^2}$-wave and $p_{x+y}$-wave occurs around $n\sim 0.4$.
\end{abstract}


\maketitle
\section{Introduction} 
Since the discovery of high-temperature superconductivity \cite{Bednorz}
the purse of identifying the microscopic mechanism for unconventional
superconductivity such as $d$-wave in cuprates or $p$-wave in ruthenates 
has been a main driving force in condensed matter physics. Even now, 
19 years after the discovery of high-temperature superconductivity,
it is an issue of hot debate whether the one-band (two-dimensional) Hubbard
model, the simplest model for electronic correlations, is becoming $d$-wave or -further away from half-filling-
$p$-wave superconducting at low temperatures.

There are indications that there is indeed 
$d_{x^2-y^2}$-wave superconductivity  close to
half-filling. \cite{ScalapinoPR} For example, 
the functional 
renormalization group (fRG)  predicts 
superconductivity for weak coupling.\cite{fRG1}
Also numerical
quantum Monte Carlo (QMC) simulations for finite size systems \cite{KurokiQMC} observe an enhancement 
of the  $d_{x^2-y^2}$-wave pairing.
A promising method for addressing these questions
is the 
dynamical cluster approximation (DCA) ,\cite{DCA} an
 extension of dynamical mean field theory (DMFT),\cite{dmft} which
also takes non-local correlations and wave-vector dependences into
account. This is essential for describing $p$- and  $d$-wave
superconductivity; DMFT itself only
allows for $s$-wave superconductivity by construction.
Such DCA calculations \cite{Maier05} find $d_{x^2-y^2}$-wave
superconductivity.
However, when solving the DCA equations numerically
by conventional
Hirsch-Fye QMC simulations \cite{HirschFye}
one is restricted to rather high 
temperatures $T$.\cite{Arita04,Maier05} Hence, a  difficult
 extrapolation to low $T$ is necessary which is further
hampered since  (numerically) exact statements
 also require an extrapolation
cluster size $N_c\!\rightarrow \!\infty$.

Our paper will  focus  on a different
parameter regime: Motivated by Sr$_2$RuO$_4$,
many authors have recently addressed 
the possibility of triplet superconductivity in the 
two-dimensional one-band Hubbard model with finite next nearest neighbor 
hopping $t'$ (the $t$-$t'$ Hubbard model) at intermediate electron densities.
The results have been controversial as regards the dominant pairing symmetry:
 fRG calculations \cite{Honerkamp01,Honerkamp04,Katanin03} concluded 
that when $t'\!>\!0.3t\!-\!0.4t$  and the Fermi level is at the van 
Hove (vH) singularity, the system becomes ferromagnetic at sufficiently 
low $T$. While a $d$-wave phase spreads next to the ferromagnetic
phase in the $U$-$t'$ diagram for vH band fillings, a $p$-wave phase exists 
when going away from the vH band filling and for sufficiently large $t'$.
Third-order  perturbation theory \cite{Nomura} showed that triplet 
superconductivity is realized even when ferromagnetic spin fluctuation are 
not dominant. On the other hand, QMC,\cite{Kuroki04} FLEX,\cite{Kuroki04} and 
DCA \cite{Arita04} calculations concluded that triplet pairing does not become 
dominant for intermediate filling even when $t'$ is as large as $0.4t$.

It should be noted that fRG,  third-order perturbation theory and 
the FLEX approximation are valid only for weak coupling; finite-size QMC is 
possible only for $U\! <\! 2t$ because of a serious negative sign 
problem;\cite{Kuroki04} and the DCA has been performed only for 
high $T$s.\cite{Arita04} 
Thus, the question concerning $p$-wave (triplet) superconductivity in 
the Hubbard model  is not conclusive yet.

In this paper, we introduce a new route to address this question,
solving the DCA equations by an extended version of the projective 
QMC (PQMC) method.\cite{Feldbacher}  This DCA(PQMC) approach mitigates 
the $T$-extrapolation problem of  DCA(QMC). We concentrate on the 
crossover between $d$- and $p$-wave instability in the intermediate
electron density range $0.3\lesssim n \lesssim0.8$ and present 
results for the paramagnetic spectral function and the dominant pairing
symmetry for $U\lesssim4t$ and $t'=0.4t$. On an equal
footing with these pair susceptibilities, we also
calculate the ferromagnetic and 
antiferromagnetic spin susceptibility of the DCA cluster.

\section{$t$-$t'$ Hubbard model} 
The $t$-$t'$ Hubbard model reads
\begin{equation}
\!H=-t\sum_{ i,j,\sigma}^{\rm N\!N}
c_{i\sigma}^{\dagger}c_{j\sigma}
+t'\sum_{ i,j ,\sigma}^{\rm N\!N\!N}
c_{i\sigma}^{\dagger}c_{j\sigma}
+U\sum_{i}n_{i\uparrow}n_{i\downarrow}.
\end{equation}
Here, $c_{i\sigma}^{\dagger}$ and $c_{i\sigma}$ create and annihilate
an electron with spin $\sigma$ on site $i$ of the two-dimensional
lattice; the first and second sum are restricted
to 
nearest neighbors (NN) and next-nearest neighbors (NNN), respectively.
In the following, all energies are given in units of $t$,
corresponding to a bandwidth of $D=8(t)$.

\section{DCA(PQMC) method} 
In the DCA,\cite{DCA}  the Brillouin zone is divided into $N_c$ patches, with
a coarse-grained Green function $\bar{G}({\bf K}_p,\omega)$
and self energy $\Sigma_c({\bf K}_p,\omega)$ for every patch $p$:
\begin{equation}
\bar{G}({\bf K}_p,\omega)=\frac{N_c}{N}\sum_{\tilde{\bf k}}
\big[G_0^{-1}({\bf K}_p+\tilde{\bf k}) -\Sigma_c({\bf K}_p,\omega)\big]^{-1}.
\label{DCAeq}
\end{equation}
Here, $G_0^{-1}$ is the non-interacting Green function
and the $\tilde{\bf k}$ summation averages over all ${\bf k}$-points of
patch $p$ which is
centered around ${\bf K}_p$; $N$ is the total number of all ${\bf k}$-points.
$\bar{G}$ and $\Sigma_c$ then define an effective cluster
of Anderson impurities which can be described by the non-interacting 
Green function 
\begin{equation}
{\cal G}_0({\bf K}_p,\omega)^{-1}=
\bar{G}({\bf K}_p,\omega)^{-1}+\Sigma_c({\bf K}_p,\omega),
\end{equation}
or its Fourier transform  ${\cal G}_0(\omega)_{{\bf X}_i{\bf X}_j}$.
This defines a cluster problem $H_c$  given by 
 ${\cal G}_0(\omega)_{{\bf X}_i{\bf X}_j}$ and a local Coulomb interaction
on every cluster site $X_i$. In the DCA, this cluster problem
has to be solved self-consistently together with
Eq.\ (\ref{DCAeq}).

Here, we introduce a new cluster solver which is based
on PQMC \cite{Feldbacher}
and particularly constructed for zero temperature.
Just as in  Ref.\ \onlinecite{Feldbacher}, 
$T=0$ expectation values  of an
arbitrary operator ${\cal O}$ are calculated as:
\begin{eqnarray}
\langle\mathcal{O}\rangle_{0}
&=&\lim_{\theta\rightarrow\infty}
\lim_{\tilde{\beta}\rightarrow\infty}
\frac
{\operatorname*{Tr}
\left[  e^{-\tilde{\beta}H_{T}}e^{-\theta H_c/2}
\mathcal{O}e^{-\theta H_c/2}\right]  }
{\operatorname*{Tr}\left[
e^{-\tilde{\beta}H_{T}}e^{-\theta H_c} \right]  }.
\label{Eq:projection}
\end{eqnarray}
where $H_{T}$ is an auxiliary Hamiltonian
 for which
we take the  cluster defined by ${\cal G}_0(\omega)_{{\bf X}_i{\bf X}_j}$
without Coulomb interaction ($U\!=\!0$).
The local one-particle
potential of  $H_{T}$  is adjusted self-consistently 
to yield the same $n$ as the interacting cluster $H_c$. At least for $N_c=1$, this
yields the same 
asymptotic behavior of ${\cal G}_0(\tau)$ and $\bar G(\tau)$
for large  $\tau$. Hence, the influence of 
breaking time translational 
symmetry at $\tau\!=\!\theta$ is reduced. This gives a smoother
Green function in the vicinity of  $\tau\!=\!\theta$,
but does not affect the  results after long enough projection, i.e.,
sufficiently away from   $\tau\!=\!\theta$.

As in  Ref.\ \onlinecite{Feldbacher},  the
 limit $\tilde{\beta}\rightarrow \infty$
can be taken analytically for this $H_T$.
Then, the interacting Green function $G$
is obtained via the same updating equations for the
auxiliary Hubbard-Stratonovich fields
as for  finite-$T$ QMC.
But the PQMC starting point is different:
 a $T\!=\!0$  Green function with
open boundary conditions
defined for  $0\leq\tau,\tau'\leq\theta=16t,18t$,
discretized  into $L=48,64$ slices for $N_c\!=\!4\times\! 4\!=16$
cluster sites.
For the measurement of physical quantities,
${\cal L}=8,10,12$ central time slices  are taken,
and the remaining ${\cal P}$ time slices on the right and
left side of the measuring interval are reserved for projection.
Typically, we performed $10^5$ to  $6\times 10^5$ QMC sweeps.
To obtain $G(i\omega)$ from  $G(\tau)$,
the maximum entropy method (MEM) is employed,
as in  Ref.\ \onlinecite{Feldbacher}.

\section{Spectral properties of the paramagnetic phase}
\begin{figure}[tb]
\includegraphics[width=8cm]{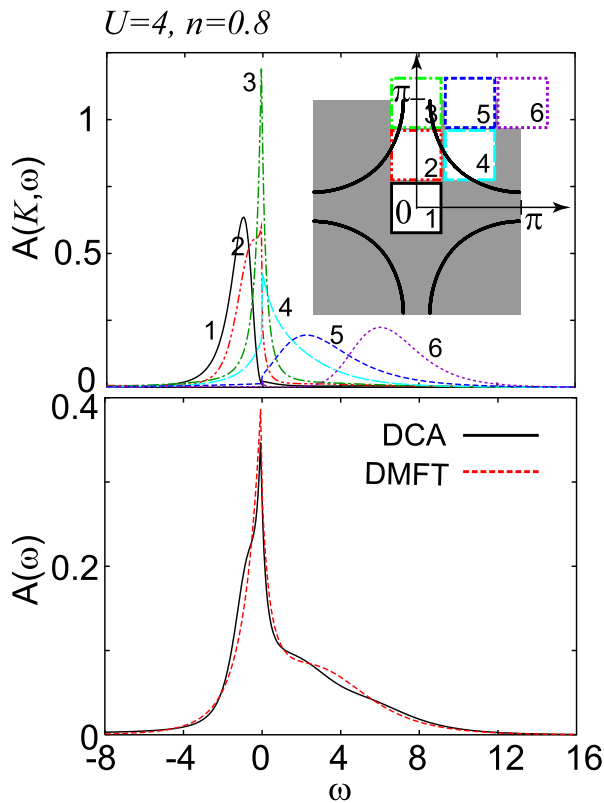}
\caption{(Color online) Top panel: Spectral 
function $A(K_p,\omega)$ for $U\!=\!4t,\! n\!=\!0.8$
and  the 6  (out of 16) irreducible patches $p$
indicated in the inset. The inset
also shows the  non-interacting Fermi surface
as a line.
 Bottom panel:
${\bf k}$-integrated spectral function $A(\omega)$
compared with that of DMFT.
}
\label{Fig:Spectrum}
\end{figure}

Now, let us turn to the DCA(PQMC) results. We performed
calculations for $n\!=\!0.3, 0.4, 0.6, 0.8$ and
$U\!=\!2t, 3.5t, 4t$. Throughout the study, we took $t'\!=\!0.4t$
and $N_c\!=\!4\!\times\! 4\!=\!16$ DCA patches.
In the upper panel of Fig.\ \ref{Fig:Spectrum}, we present
 the  one-particle spectral function
 $A({\bf K}_p,\omega)=-1/\pi\, {\rm Im} \bar{G}({\bf K}_p,\omega)$, 
i.e., the averaged spectrum over the
six irreducible patches $p$ of the
Brillouin zone, as indicated in the inset.
Patches 2, 3, and 4
have contributions at the Fermi surface and
hence a contribution at $\omega=0$. Thereby,
$A({\bf K}_3,\omega)$ has a particular sharp peak
because the vH singularity is located at $(\pi,0)$ and 
$(0,\pi)$.
This sharp peak 
also reflects in the fully ${\bf k}$-integrated
$A(\omega)=\sum_p A({\bf K}_p,\omega)$, 
shown in the bottom panel of Fig.\ \ref{Fig:Spectrum}.
Let us note that the quantitative and qualitative features are 
very similar to those of DMFT (dashed line);
there is no pseudo gap for $t'\!=\!0.4, n\!=\!0.8, U\!=\!4t$ in our $T\!=\!0$ DCA(PQMC) calculation.

\section{Calculation of cluster susceptibilities}
To discuss the magnetic and pairing instabilities,
we first calculate the corresponding 
susceptibilities
\begin{eqnarray}
\chi^{\rm spin}({\bf Q},\tau_1-\tau_2)&=&\sum_{{\bf X}_1,{\bf X}_2}
{\rm e}^{{\rm i}{\bf Q}({\bf X}_1-{\bf X}_2)}\nonumber \\
&&\!\!\!\!\!\!\!\!\!\!\!\!\!\!\!\!\!\!\!\!\!\!\!\!\!\!\!\!\langle {\rm T}_\tau
c_{{\bf X}_1\uparrow}^{\dagger}(\tau_1)
c_{{\bf X}_1\downarrow}^{\phantom{\dagger}}(\tau_1)
c_{{\bf X}_2\downarrow}^{\dagger}(\tau_2)
c_{{\bf X}_2\uparrow}^{\phantom{\dagger}}(\tau_2)
\rangle,\label{sus1}\\
\chi^{\rm pair}(\tau_1-\tau_2)&=&\sum_{{\bf X}_1,{\bf X}_2}
g({\bf K}_1)g({\bf K}_2)\nonumber \\
&&\!\!\!\!\!\!\!\!\!\!\!\!\!\!\!\!\!\!\!\!\!\!\!\!\!\!\!\!\!\!\!\!\langle {\rm T}_\tau
c_{{\bf K}_1\uparrow}^{\phantom{\dagger}}(\tau_1)
c_{-{\bf K}_1\downarrow}^{\phantom{\dagger}}(\tau_1)
c_{-{\bf K}_2\downarrow}^{\dagger}(\tau_2)
c_{{\bf K}_2\uparrow}^{\dagger}(\tau_2)
\rangle
\label{sus2}
\end{eqnarray}
for  the $4\!\times 4\!$ cluster at self-consistency.
Here, $c_{{\bf K}\sigma}\!=\!\sum_{{\bf X}}c_{{\bf X}\sigma}\exp(i {\bf KX})$ and
$g({\bf K})$ is the form factor, i.e.,
\begin{equation}
g({\bf K})\!=\! \left\{
\begin{array}{l l}1 & \mbox{for $s$-wave,}\\
\sqrt{2}\sin(K_x) & \mbox{for $p_x$ wave,}\\
\sqrt{2}\sin(K_x+K_y) & \mbox{for $p_{x+y}$ wave,}\\
\cos(K_x)\!-\!\cos(K_y) & \mbox{for $d_{x^2-y^2}$ wave.}
\end{array}\right.
\end{equation}
Second, we calculate ${\rm Im}\chi(\omega)$ from $\chi(\tau)$ 
by MEM and from this obtain the static susceptibility via
the
Kramers-Kronig relation 
\begin{equation}
\chi(\omega=0)\!=\!\int {\rm Im}\chi(\omega)/\omega d\omega.
\end{equation}

In DCA, the susceptibilities of 
the Hubbard model are calculated from the above cluster
$\chi$
by solving the Bethe-Salpeter equation.\cite{DCA} 
However, to this end it would be necessary 
to calculate susceptibilities for two $\tau$'s  instead of Eqs.\   (\ref{sus1}) 
and (\ref{sus2}) which would tremendously increase the numerical effort.
Hence, we look at the
cluster susceptibilities Eqs.\ (\ref{sus1}) 
and (\ref{sus2})  for simplicity.
We can expect that this quantity already captures the essential features 
of the competition between ferromagnetic and antiferromagnetic 
spin fluctuation, or that between superconductivity with
different symmetries.

\begin{figure}[tb]
\includegraphics[width=12cm]{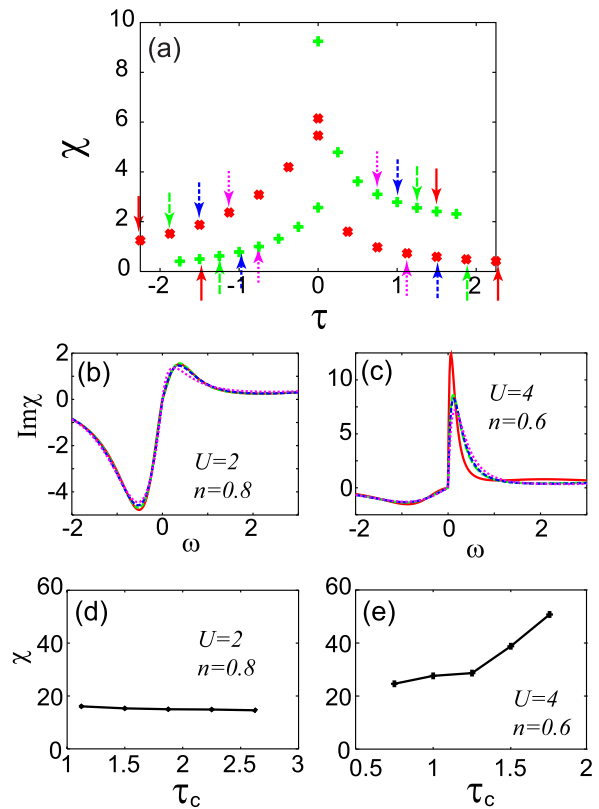}
\caption{(Color online) 
a) Susceptibilities for $d_{x^2-y^2}$-wave pairing in  $\tau$-space
 for $U\!=\!2t, n\!=\!0.8$ (crosses) and $U\!=\!4t, n\!=\!0.6$ (circles). 
The
arrows indicate different cutoffs $\tau_c$ 
up to which  $\chi(\tau)$ information
was taken into account
in MEM for obtaining the ${\rm Im}\chi(\omega)$ 
in part b) and c).
Dotted, dashed, long dashed, and solid line 
are 
for the corresponding $\tau_c$-$\downarrow$ of 
 the main figure.
For $U\!=\!2t, n\!=\!0.8$ (panel b),
the result is basically  $\tau_c$-independent;
but for  $U\!=\!4t, n\!=\!0.6$ (panel c)
the peak at small $\omega$ is largely enhanced
if $\tau_c$ is increased.
Consequently, the static susceptibility 
$\chi(\omega=0)$ is basically
$\tau_c$-independent in panel d)
and strongly $\tau_c$-dependent in panel e).
In the following, we indicate the
problematic cases as in panel e)  by an arrow.
}
\label{Fig:chi}
\end{figure}

Let us mention one more technical point: 
The decay of $\chi$ in $\tau$-space
can be underestimated because of a 
finite  number of projection time slices
${\cal P}$.
In this case, the MEM calculation of 
${\rm Im} \chi(\omega)$ depends on up to which
$\tau_c$ value the MEM input  $\chi(\tau)$  
is considered, see Fig.\ (\ref{Fig:chi}).
Whenever we encounter a strong
 cutoff ($\tau_c$-)dependence like in Fig.\ \ref{Fig:chi} (right inset)
we take a short cutoff $\tau_c$  (only four $\tau$ points).
With such a short cutoff $\tau_c$ which is very close
to the asymptotical behavior  $\tau_c\rightarrow 0$,
we certainly do not overestimate   $\chi(0)$.
However, if the  weak decay  of $\chi$ in $\tau$-space
is not due to the finite ${\cal P}$ but
due to physics, 
we {\em underestimate} $\chi(0)$
this way.
In Figs.\ \ref{Fig:magsus} and \ref{Fig:pairsus}, 
these more problematic data points, 
where we possibly  {\em underestimate} $\chi(0)$,
are indicated by an arrow.

\section{Instabilities of the paramagnetic phase}
\begin{figure}[tb]

\vspace{.3cm}

\includegraphics[width=8.5cm]{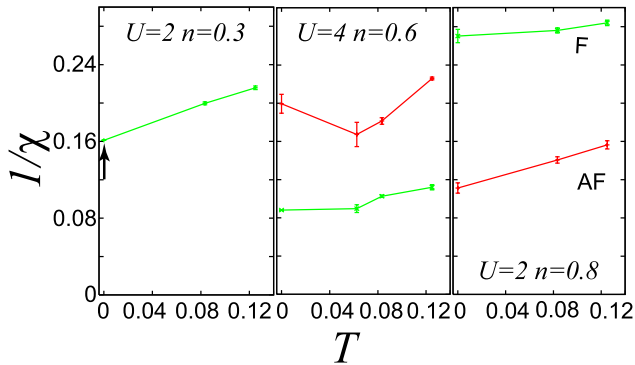}
\caption{(Color online) 
Inverse magnetic susceptibilities as a function of $T$ for
$U\!=\!2t,n\!=\!0.3$ (left), $U\!=\!4t,n\!=\!0.6$ (middle), and 
$U\!=\!2t,n\!=\!0.8$ (right). The arrows indicate those
data points  where $\chi$ is  possibly underestimated
($1/\chi$ overestimated),
as discussed in Fig.\ \ref{Fig:chi}. 
\label{Fig:magsus}}
\end{figure}

Let us finally turn to  the physical results, starting
with the (inverse) ferromagnetic and antiferromagnetic  
susceptibilities plotted in Fig.\ \ref{Fig:magsus}
as a function of $T$ for three exemplary parameter values;
finite-$T$ data have been obtained by conventional DCA(PQMC).
First of all, let us emphasize the obvious:
DCA(PQMC) directly provides well 
for the $T\!=\!0$ susceptibilities 
{\em without} extrapolation
-in very contrast to the finite-$T$ DCA(QMC)
for which it is difficult to forecast/extrapolate
the low-$T$ behavior.
Turning to the physical results, we 
see that the ferromagnetic spin susceptibility
becomes larger than the antiferromagnetic one
for $n\le 0.6$. Note that the vH filling is at $\sim 0.6$ for $t'=0.4$. 
In fact, the fRG studies \cite{Honerkamp04,Katanin03} found a strong 
ferromagnetic instability for $t'>0.3\sim 0.4$ at vH band fillings. Thus,
the qualitative tendency of our results and those of the fRG 
studies are consistent.

\begin{figure}[tb]

\vspace{.3cm}

\includegraphics[width=8.5cm]{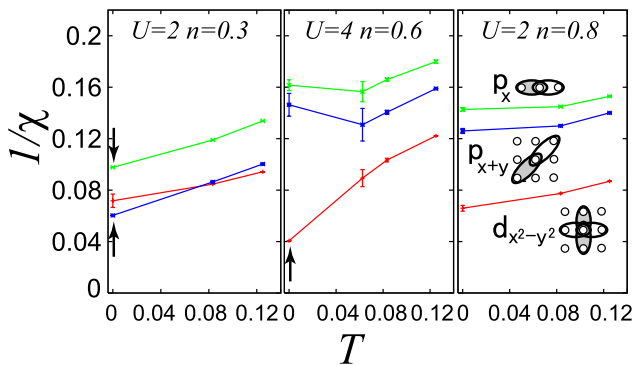}
\caption{(Color online) 
Same as  Fig.\ \ref{Fig:magsus}, but now
for  $p_x$, $p_{x+y}$, and  $d_{x^2-y^2}$-wave
 pairing as indicated
in the right panel.
\label{Fig:pairsus}}
\end{figure}

Next, let us discuss superconductivity. Fig.\ \ref{Fig:pairsus}
shows the inverse static
pair susceptibilities as a function of $T$.
First of all, we find that $d_{x^2-y^2}$-wave ($p_{x+y}$-wave) 
is dominant among singlet (triplet) pairings
for all $(U,n)$'s considered.
For $n\!=\!0.8$,  $d_{x^2-y^2}$-wave dominates over triplet
pairings as  expected. 
When we reduce $n$ to 0.6, antiferromagnetic spin fluctuation
become weak, see Fig.\ \ref{Fig:magsus}. Nonetheless, the 
$d_{x^2-y^2}$-wave  instability is still larger than
those for triplet pairings. This is consistent
with finite temperature DCA 
calculation of Ref.\ \onlinecite{Arita04} for $U\!=\!3t,n\!=\!0.67$, and suggests $d_{x^2-y^2}$-wave ordering for low enough $T$ -even
for far away from half-filling. 
These findings are also in accord  with QMC calculations
for a finite-size Hubbard model,\cite{Kuroki04}
showing  a dominant $d_{x^2-y^2}$-wave instability
for $t'\!\sim\! 0.4, n\!\sim\! 0.67$, and $U\!=\!2t$.
The fRG study Ref.\ \onlinecite{Katanin03} reports also
a ferromagnetic phase for 
$t'\! =\! 0.4,n\! \sim\!  0.6$, and $U\! >\! t$, while
there is possibly a $d$-wave phase for $U<t$. 
On the other hand, third ordered perturbation 
calculations \cite{Nomura} have shown that triplet 
superconductivity becomes dominant even for weak coupling.
We find such triplet pairing only for smaller $n$'s,
while the detailed structure of ${\bf k}$-dependence of
the vertex corrections may be essential which is possibly
smeared out by the coarse-graining of DCA. 

\section{Discussion and Outlook}
\begin{figure}[tb]
\vspace{.5cm}
\includegraphics[width=8.5cm]{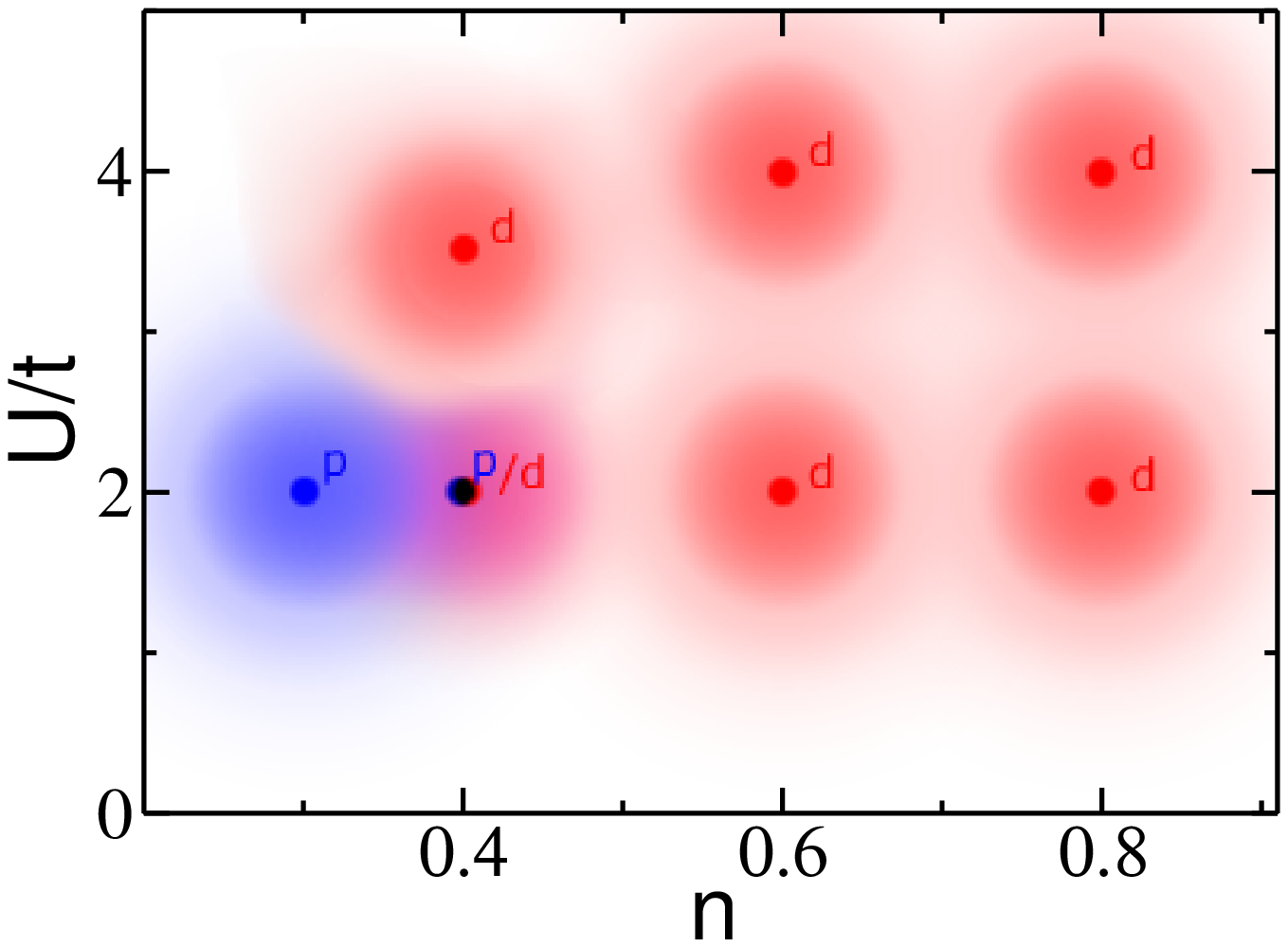}
\caption{(Color online) 
$U$-$n$ diagram  showing the dominant pairing symmetry; for 
$n\!=\!0.4$, $U\!=\!2t$ it is too close to call which
instability is strongest.}
\label{Fig:phasdiag}
\end{figure}

Fig.\ \ref{Fig:phasdiag} summarizes our main finding, i.e.,
the dominant superconducting susceptibility on the
cluster.
We find that $d_{x^2-y^2}$-wave pairing is dominant 
in an unexpectedly  broad range of fillings $n$, whereas
  $p_{x+y}$-wave prevails only for small $n$.
The transition between  $p_{x+y}$-wave and $d_{x^2-y^2}$-wave 
is at about $n\!\sim\!0.4$.

The $t$-$t'$ Hubbard model with $t'\!\sim\! 0.4$ and $n\!\sim \!2/3$ has been 
considered to be a good model for the $\gamma$-band
of Sr$_2$RuO$_4$.\cite{Nomura}
Our DCA(PQMC) results indicate however 
that there is no triplet superconductivity 
for these $n$ and $t'$ values, at least  for
the moderately strong values of $U$
we studied and which are e.g.\
considered in 
local density approximation (LDA)+DMFT calculations.\cite{Liebsch}

Hence our results suggest that 
a nearest-neighbor Coulomb interaction $V$, as suggested in 
Ref.\ \cite{Arita04}, or
a multi-orbital model is
 necessary for an appropriate description
of triplet pairing in Sr$_2$RuO$_4$
below $T_c\!\sim\! 1\,$K.
In contrast to the  standard finite-$T$ DCA(QMC), 
we can expect that DCA(PQMC) 
will open the door to such 
low temperatures  -even for calculations with 
orbital realism. 
Similarly, we can hope that  DCA(PQMC) 
will allow for more definite statements
concerning superconductivity 
in the two-dimensional Hubbard model in the future.
To this end, still a $N_c\!\rightarrow\!\infty$
extrapolation and the calculation of lattice susceptibilities
is necessary. But at least the
problematic  extrapolation to
low temperatures is mitigated; PQMC is the direct route
 to $T\!=\!0$.
 
We acknowledge discussions with M.\ Feldbacher and A.\ A.\ Katanin, and thank the Alexander von Humboldt 
foundation (RA) and the Emmy Noether program of 
the Deutsche Forschungsgemeinschaft (KH) for financial support.

\end{document}